\documentclass[aps,prapplied,%
reprint,%
a4paper,superscriptaddress,floatfix,amsmath,amssymb,amsfonts,noshowpacs]{revtex4-2}
\pdfoutput=1
\usepackage{newtxtext,newtxmath}
\usepackage[utf8]{inputenx}
\usepackage{graphicx}
\usepackage[dvipsnames]{xcolor}
\usepackage{soul} 
\usepackage[textwidth=17.5cm,textheight=23.5cm,verbose,pdftex]{geometry}
\usepackage{floatrow}
\usepackage[pdftex]{hyperref}

\hypersetup{pdfauthor={Thomas Auzelle}, bookmarksnumbered=true, pdftitle={Interface recombination in Ga- and N-polar GaN/(Al,Ga)N quantum wells grown by molecular beam epitaxy}, colorlinks, citecolor=blue, linkcolor=blue, urlcolor=blue}

\usepackage{gensymb}
\usepackage{graphicx}

\newcommand{\etal}[1]{\textit{et al.}\,[\onlinecite{#1}]}
\newcommand{\ie}{\textit{i.e.}}
\newcommand{\eg}{\textit{e.g.}}
\newcommand{\degC}{~$\celsius$}

\newcommand{\insitu}{\textit{in~situ}}

\begin{document}
	\title{Interface recombination in Ga- and N-polar GaN/(Al,Ga)N quantum wells grown by \\molecular beam epitaxy}

\author{T. Auzelle} \email[Electronic mail: ]{auzelle@pdi-berlin.de}
\affiliation{Paul-Drude-Institut für Festkörperelektronik, Leibniz-Institut im Forschungsverbund Berlin e.\,V., Hausvogteiplatz 5--7, 10117 Berlin, Germany}

\author{C. Sinito}
\affiliation{Paul-Drude-Institut für Festkörperelektronik, Leibniz-Institut im Forschungsverbund Berlin e.\,V., Hausvogteiplatz 5--7, 10117 Berlin, Germany}
\affiliation{Present address: WILCO AG, Rigackerstrasse 11, 5610 Wohlen, Switzerland}

\author{J. Lähnemann}
\affiliation{Paul-Drude-Institut für Festkörperelektronik, Leibniz-Institut im Forschungsverbund Berlin e.\,V., Hausvogteiplatz 5--7, 10117 Berlin, Germany}

\author{G. Gao}
\affiliation{Paul-Drude-Institut für Festkörperelektronik, Leibniz-Institut im Forschungsverbund Berlin e.\,V., Hausvogteiplatz 5--7, 10117 Berlin, Germany}
\affiliation{Present address: Materials Science and Nano-engineering Department, Rice University, Houston, Texas 77005, United States}

\author{T. Flissikowski}
\affiliation{Paul-Drude-Institut für Festkörperelektronik, Leibniz-Institut im Forschungsverbund Berlin e.\,V., Hausvogteiplatz 5--7, 10117 Berlin, Germany}

\author{A. Trampert}
\affiliation{Paul-Drude-Institut für Festkörperelektronik, Leibniz-Institut im Forschungsverbund Berlin e.\,V., Hausvogteiplatz 5--7, 10117 Berlin, Germany}

\author{S. Fern\'andez-Garrido}
\affiliation{Paul-Drude-Institut für Festkörperelektronik, Leibniz-Institut im Forschungsverbund Berlin e.\,V., Hausvogteiplatz 5--7, 10117 Berlin, Germany}
\affiliation{Grupo de Electr\'onica y Semiconductores, Dpto. F\'isica Aplicada, Universidad Aut\'onoma de Madrid, 28049 Madrid, Spain}

\author{O. Brandt}
\affiliation{Paul-Drude-Institut für Festkörperelektronik, Leibniz-Institut im Forschungsverbund Berlin e.\,V., Hausvogteiplatz 5--7, 10117 Berlin, Germany}

\begin{abstract}
	We explore and systematically compare the morphological, structural and optical properties of GaN/(Al,Ga)N multiple quantum wells (MQWs) grown by plasma-assisted molecular beam epitaxy (PA-MBE) on freestanding GaN$(0001)$ and GaN$(000\bar{1})$ substrates. Samples of different polarity are found to be comparable in terms of their morphological and structural perfection and exhibit essentially identical quantum well widths and Al content. Regardless of the crystal orientation, the exciton decay in the MQWs at $10$~K is dominantly radiative and the photoluminescence (PL) energy follows the quantum confined Stark effect (QCSE) for different quantum well widths. A prominent free-to-bound transition involving interface shallow donors is, however, visible for the N-polar MQWs. At room-temperature, in contrast, the exciton decay in all samples is dominated by nonradiative recombination taking place at point defects, presumably Ca or V$_N$ located at the bottom QW interface. Remarkably, the N-polar MQWs exhibit a higher PL intensity and longer decay times than the Ga-polar MQWs at room-temperature. This improved internal quantum efficiency is attributed to the beneficial orientation of the internal electric field that effectively reduces the capture rate of minority carriers by interface trap states. 
\end{abstract}


\maketitle
\section{INTRODUCTION}\label{sec:introduction}
The adverse effect of dislocations on the internal quantum efficiency of GaN/(Al,Ga)N heterostructures can be essentially resolved by reducing the dislocation density below $5\times10^7$\,cm$^{-2}$ \cite{Karpov2002,Ban2011}, a regime directly attainable for growth on freestanding GaN or AlN substrates. Nonradiative processes in these heterostructures are thus mediated by nonradiative point defects \cite{Sampath2010,Chichibu2018}, whose density must be lowered to reach high luminous efficiencies. In this regard, optimization of the III/V ratio or Si doping have been reported to enhance the internal quantum efficiency of metal polar (Al,Ga)N quantum wells (QWs) grown by metal organic chemical vapor epitaxy (MOCVD) \cite{Murotani2012,Bryan2015}. Yet, establishing other strategies to further decrease the density of point defects in GaN/(Al,Ga)N heterostructures would certainly benefit to the performance of nitride-based ultraviolet light emitters \cite{Kneissl2019}, transistors \cite{Eller2013,Huber2015,Hu2019,Dangeabf1388}, and future devices. For the epitaxial growth of semiconductors, it is well known that the crystal orientation influences not only the surface morphology \cite{Lymperakis2009,Sawicka2011,Shao2013,Sawicka2017,Tatarczack2021}, but also the incorporation of impurities and the formation of nonradiative point defects \cite{Brandt1995,Ptak2001,Tuomisto2005}. For instance, a dramatic case are (In,Ga)N/GaN QWs grown by plasma-assisted molecular beam epitaxy (PA-MBE), which may not exhibit any detectable luminescence even at $10$~K if grown N-polar [\ie, on GaN$(000\bar{1})$] instead of Ga-polar [\ie, on GaN$(0001)$] \cite{Cheze2013b,Garrido2016}. The origin of this phenomenon is related to the presence of nonradiative defects at the QW/barrier interface. A potential candidate for this as yet unknown point defect is the impurity Ca \cite{Shen2017,Cheze2018}, which can contaminate the substrate surface during wet cleaning and gradually incorporates in the overgrown nitride layers \cite{Young2016}.

%

Growth of GaN/(Al,Ga)N heterostructures on various crystal orientations has been already reported by MBE \cite{Sarigiannidou2006,Natali2009,Smalc2011,Mazumder2012,Shao2013b,Fireman2017} and MOCVD \cite{Kuokstis2002,Vashaei2010, Keller2012,Banal2014,Rosales2014}, but a systematic investigation of the impact of crystal orientation on the incorporation rate of nonradiative point defects is still lacking. A particular interest lies in the $(000\bar{1})$ orientation since it exhibits internal electric fields of opposite direction compared to the classical $(0001)$ orientation, which is seen as favorable for improving the efficiency of high-frequency transistors \cite{Wong2013}, light emitting diodes \cite{Akyol2011,Han2012,Dong2012,Feng2015} and solar cells \cite{Li2010}. Last, the $(000\bar{1})$ orientation is currently available in the form of freestanding GaN substrates at similar costs than the classical $(0001)$ ones and is also the usual growth direction of dislocation-free GaN nanowires obtained by self-assembly \cite{Zuniga-Perez2016}. 

In this work, we systematically compare the morphological, structural and optical properties of $(0001)$ and $(000\bar{1})$ GaN/(Al,Ga)N multiple quantum wells (MQWs) grown by PA-MBE and discuss the observed differences in terms of point defect incorporation at the bottom QW interface.
As a prerequisite, the samples of different polarity are first evidenced to be comparable in terms of their morphological and structural perfection. Low concentrations of impurities are found, with the exception of Ca, whose concentration is seen to be high in the Ga-polar samples, ten times higher than for the N-polar samples. For all the MQWs, the low temperature photoluminescence (PL) is showing a predominantly radiative behavior, whereas the room-temperature PL is dominated by nonradiative interface recombination. However, the N-polar MQWs exhibit an improved internal quantum efficiency at room-temperature in comparison to the Ga-polar MQWs, an effect that originates from the opposite direction of the internal electric field.

\section{EXPERIMENTS AND METHODS}\label{sec:experimental_details}
\begin{figure}
\includegraphics[width = .8\linewidth]{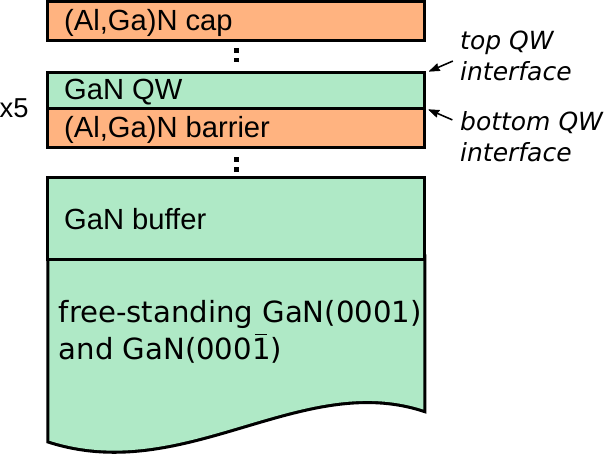}%
\caption{Nominal structure of the series of GaN/(Al,Ga)N MQWs under investigation
\label{fig:MQWstruct}}
\end{figure}

The MQWs samples under investigation consist of the layer sequence sketched in Fig.~\ref{fig:MQWstruct}.
They are grown in a PA-MBE setup equipped with a Veeco plasma cell to generate active N and effusion cells for Ga and Al. The atomic fluxes are calibrated by growing thick GaN(0001) and AlN(0001) layers as described in Ref.~\cite{Heying2000}.
Free-standing GaN is chosen as substrate to nominally provide equivalent structural quality for Ga- and N-polar samples. The substrates were purchased from Suzhou Nanowin Science and Technology, with a nominal threading dislocation density below $5 \times 10^{6}$~cm$^{-2}$ and a $0.35 \pm 0.15^\circ$ miscut toward the $[1\bar{1}00]$ direction. 
The in-house substrate preparation consists in a $5$~min dip in a $5\%$ HF solution before mounting them with In on a 2" GaN/Al$_2$O$_3$ template backside coated with Ti. They are next degassed in vacuum for $20$ min at $300$\degC{} and \insitu{} cleaned right before growth by a sequence of Ga deposition/desorption steps at $\approx 600/750$\degC{}. This preparation is aimed at reducing the amount of residues on the GaN surface to prevent the formation of extended defects at the regrowth interface \cite{Storm2012,Storm2015,Storm2016,Blanchard2018}.
Next, a $\approx 600$~nm thick buffer of GaN is grown in order to bury the residual impurities located on the substrate surface. Nominal atomic fluxes of $5.7 \times 10^{14}$~s$^{-1}$\,cm$^{-2}$ for N and $8.5 \times 10^{14}$~s$^{-1}$\,cm$^{-2}$ for Ga are used, corresponding to a Ga surplus of $50\%$ relative to the N flux, which should ensure a smooth growth for both polarities \cite{Cheze2013a}.
The growth is N limited resulting in a growth rate of $7.7$\,nm/min. The substrate temperature is constantly tuned to set the growth conditions at the boundary below which Ga will start accumulating on the surface as droplets \cite{Heying2000,Okamura2014}. This is checked by monitoring the intensity change of the reflection high-energy diffraction pattern, which dims in the presence of metal droplets. This way, the growth conditions are optimized for each sample, independently of their polarity and of their thermal contact with the substrate holder. Based on the growth diagrams of Refs. \cite{Heying2000,Okamura2014}, the effective substrate temperature amounts to ($700\pm 10$) and ($715 \pm 10$)\degC{} for the Ga-polar and N-polar samples, respectively. The difference results from the slightly larger activation energy for desorption of Ga adatoms on the GaN$(000\bar{1})$ surface ($3.1$ -- $3.6$\,eV \cite{Koblmuller2004,Okamura2014}) compared to the GaN$(0001)$ one ($2.8$ -- $3.2$\,eV \cite{Hacke1996,Koblmuller2004}).
Prior to the first (Al,Ga)N barrier growth, the substrate temperature is nominally fixed and the Ga surplus relative to the N flux is reduced to $30\%$. This is to ensure the absence of metal accumulation when the Al flux is additionally provided for the (Al,Ga)N barrier growth. Then, the heterostructure is grown only by sequentially opening and closing the Al cell shutter, using a nominal atomic flux of $0.8 \times 10^{14}$~s$^{-1}$\,cm$^{-2}$. No growth interruptions have been performed to smooth the interfaces.  The low growth temperature used here should prohibit exchanges between Ga and Al atoms during capping of the GaN QWs by the (Al,Ga)N barrier \cite{Gogneau2004a,Kuchuk_2014}. Hence, the growth rate of the GaN quantum wells and (Al,Ga)N barriers is limited by the N flux and the Al content within the (Al,Ga)N barriers corresponds to the Al/N ratio \cite{Iliopoulos2002}. 
Two series of Ga- and N-polar MQWs are grown, by varying the thickness of the GaN QWs. Their structural parameters are gathered in Table~\ref{table:sample_dimensions}. Widths are indicated in nm and in monolayers (ML), where a GaN ML amounts to $0.259$~nm and an Al$_{0.1}$Ga$_{0.9}$N ML $0.258$~nm. To ensure comparable growth conditions, pairs of Ga- and N-polar samples with similar structures are grown the same day, in two subsequent runs.

\begin{table*}
	\caption{Structural properties of the investigated samples as deduced from the X-ray analysis. The widths are denoted both in monolayer (ML) and in nm and have an uncertainty of $1$\,ML related to the inhomogeneity over the sample. \label{table:sample_dimensions}}
	\begin{tabular}{ccccc}
		Polarity 	& GaN QWs width & (Al,Ga)N barrier width & (Al,Ga)N capping layer width & Al content \\\hline\hline
		Ga				& $10$~ML ($2.59$~nm) & $30$~ML ($7.75$~nm) & $126$~ML ($32.55$~nm) & $8.9\%$ \\
		N				& $10$~ML ($2.59$~nm) & $30$~ML ($7.75$~nm) & $126$~ML ($32.55$~nm) & $8.9\%$ \\
		Ga				& $17$~ML ($4.41$~nm) & $30$~ML ($7.75$~nm) & $126$~ML ($32.55$~nm) & $8.9\%$ \\
		N				& $17$~ML ($4.41$~nm) & $29$~ML ($7.49$~nm) & $124$~ML ($32.03$~nm) & $8.7\%$ \\
		Ga				& $23$~ML ($5.96$~nm) & $30$~ML ($7.75$~nm) & $126$~ML ($32.55$~nm) & $8.8\%$ \\
		N				& $22$~ML ($5.70$~nm) & $29$~ML ($7.49$~nm) & $123$~ML ($31.76$~nm) & $10.2\%$ \\
	\end{tabular}
\end{table*}

Once grown, the samples are structurally and morphologically characterized by atomic force microscopy (AFM) using the PeakForce Tapping mode of a Bruker Dimension Edge and X-ray diffraction (XRD) carried out in a PANalytical X'Pert Pro diffractometer equipped with a Ge(220) hybrid monochromator, using the Cu$_{K\alpha_{1}}$ radiation (wavelength $\lambda=1.54056$~\AA). Symmetric $\omega - 2\theta$ scans across the 0002 GaN Bragg reflection are carried out with an open detector (without crystal analyzer or slit) and the experimental diffraction scans are simulated using a dynamical X-ray diffraction model \cite{Brandt2002}. Transmission electron microscopy (TEM) is additionally performed on a pair of Ga- and N-polar samples to get a local insight into the interface structure of the heterostructure. Cross-sections are prepared by mechanical grinding and Ar-ion beam milling down to electron transparency. Measurements are performed at $200$~kV in a Jeol 2100F field emission microscope, equipped with a Gatan Ultra Scan 4000 CCD. 

Impurity concentrations are measured by secondary ion mass spectrometry (SIMS) measurements performed by RTG Mikroanalyse GmbH Berlin. O and C atoms are measured in a Cameca IMS 4s setup using a primary Cs$^+$ ion source. B and Ca atoms are measured in a Cameca IMS 4fe6 setup using a primary O$_2^+$ ion source. Quantification of the SIMS signal is done by measuring ion implanted standards. Care was taken to check that the measured $^{40}$Ca signal does not include any contributions from compounds such as $^{28}$Si$^{12}$C and $^{24}$Mg$^{16}$O.

Continuous-wave photoluminescence (cw-PL) spectroscopy is performed with the $325$~nm line of an He-Cd laser and the $244$~nm line of a frequency-doubled Ar ion laser. The PL signal is collected in backscatter geometry and dispersed by an $80$~cm monochromator and detected by a charge-coupled-device (CCD) detector cooled with liquid N. The measurements are corrected for the spectral response of the setups. 
Time-resolved photoluminescence (TR-PL) is carried out using the second harmonic of fs pulses obtained from an optical parametric oscillator pumped by a Ti:sapphire laser. The emission wavelength and the repetition rate are $325$~nm and $76$~MHz, respectively, and the energy fluence per pulse is maintained below $0.2$~$\mu$J/cm$^2$. The transient emission was spectrally dispersed by a monochromator and detected by a streak camera. Both for cw-PL and TR-PL, the samples are mounted together on a cold-finger cryostat allowing measurements at $10$~K.

The band profiles and wave functions are computed using a 1D Schr\"odinger-Poisson solver \cite{Snider}, using similar parameters as in \cite{Garrido2016}.

\section{RESULTS and DISCUSSION}\label{sec:results}
\subsection{Morphological and structural characterization}\label{subsec:morphology}
\begin{figure}
\includegraphics[width = \linewidth]{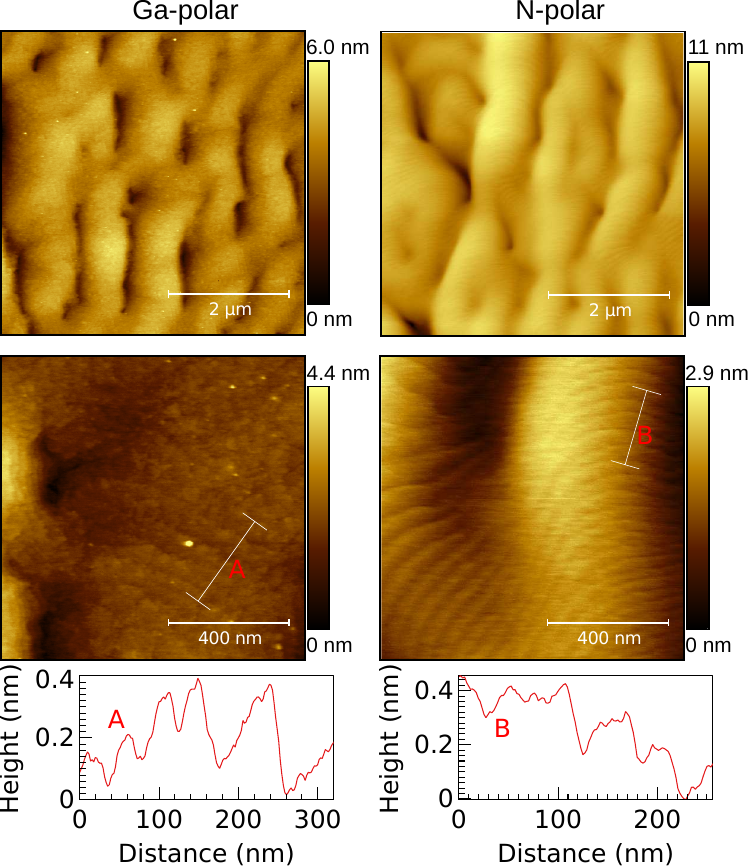}%
\caption{Representative AFM images of the pair of Ga-polar and N-polar $10$~ML GaN/(Al,Ga)N MQWs. The linescans A and B show atomic steps with a step height of 1--2 ML.
\label{fig:AFM}}
\end{figure}

Fig.~\ref{fig:AFM} shows representative AFM images from the Ga- and N-polar $10$~ML MQWs. Both samples feature a smooth surface with a root-mean-square (rms) roughness below $1.5$\,nm. Hillocks related to mixed type dislocations \cite{Heying1999} are not visible here, in agreement with the nominal substrate density of threading dislocations ($\approx 5\times 10^{6}$~cm$^{-2}$). A step-flow growth mode is confirmed by the presence of clear atomic steps with a step height equal to $1$--$2$\,ML. However, a pronounced meandering of the terrace edges is observed, resulting in few nanometer deep trenches, which causes the rms roughness value to exceed $1$\,nm. This finger-like morphology typically arises during the epitaxy of N-polar nitride layers in metal-rich conditions \cite{Cheze2013a,Cho2016,Turski2019,Diez2020} and also for the growth of Ga-polar GaN layers in specific conditions (\eg\ at $780$\degC{} with a large III/V ratio of $1.6$ \cite{Koblmuller2010}). The meanders form as a result of a large Ehrlich-Schwöbel barrier at the kinks of the step edges that makes adatom diffusion along step edges asymmetric and favors kink bunching \cite{Misbah2010}. 

\begin{figure}
\includegraphics[width = \linewidth]{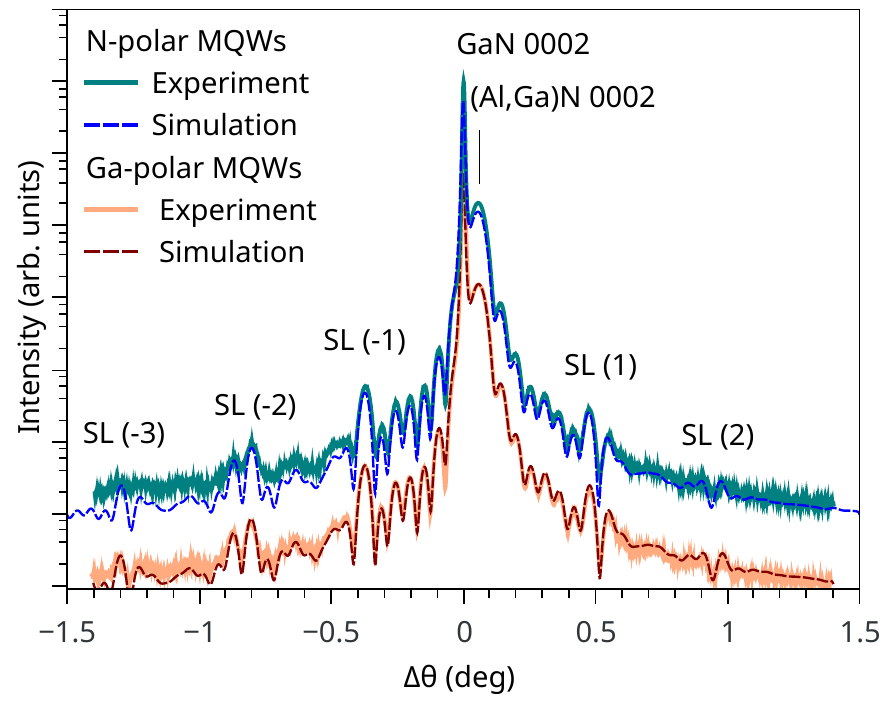}%
\caption{Representative symmetric $\omega - 2\theta$ scans across the 0002 GaN Bragg reflection of the pair of Ga- and N-polar $10$~ML GaN/(Al,Ga)N MQWs. The simulated diffraction profiles for GaN/(Al,Ga)N MQWs resulting in the optimum fit of the experimental data are superposed. Superlattice satellites (SL) are labelled and the diffraction profiles are vertically shifted for clarity.
\label{fig:xray}}
\end{figure}

Fig.~\ref{fig:xray} displays representative symmetric $\omega - 2\theta$ scans acquired across the 0002 GaN Bragg reflection with an open detector, thus resembling a rocking curve. Superlattice satellites and pronounced interference fringes related to the MQW structure are equally visible for both sample polarities. The entire profiles are in excellent agreement with those expected theoretically despite the absence of a crystal analyzer. The minimal broadening and damping is attributed to the absence of dislocation-induced strain \cite{Kaganer2015} and to an excellent spatial uniformity of the heterostructure. We note that in this diffraction configuration, the effect of interface roughness (with an rms value $\leq 2$\,ML) on the diffraction pattern is very small. The experimental diffraction profiles are fitted by simulating GaN/(Al,Ga)N MQWs following two constraints: (1) only complete monolayers are considered and (2) the atomic fluxes are assumed constant during the full heterostructure growth (\ie, equal widths for all the quantum wells of a single heterostructure). Optimum fits for the diffraction profiles of the pair of the $10$~ML MQWs are shown in Fig.~\ref{fig:xray}. The structural parameters obtained from the optimum fits of all the samples are provided in Table~\ref{table:sample_dimensions}. The error bars are set to $1$~monolayer to account for the fit accuracy and the dispersion within each sample. The N-polar heterostructures are seen on average slightly thinner than their Ga-polar counterparts, in agreement with the observations of Okamura \etal{Okamura2014}, who have reported a lower efficiency for the incorporation of N adatoms on the GaN$(000\bar{1})$ surface compared to the GaN$(0001)$ one. Beside these small differences, the Ga- and N-polar samples appear structurally equivalent. 

\begin{figure}
\includegraphics[width = \linewidth]{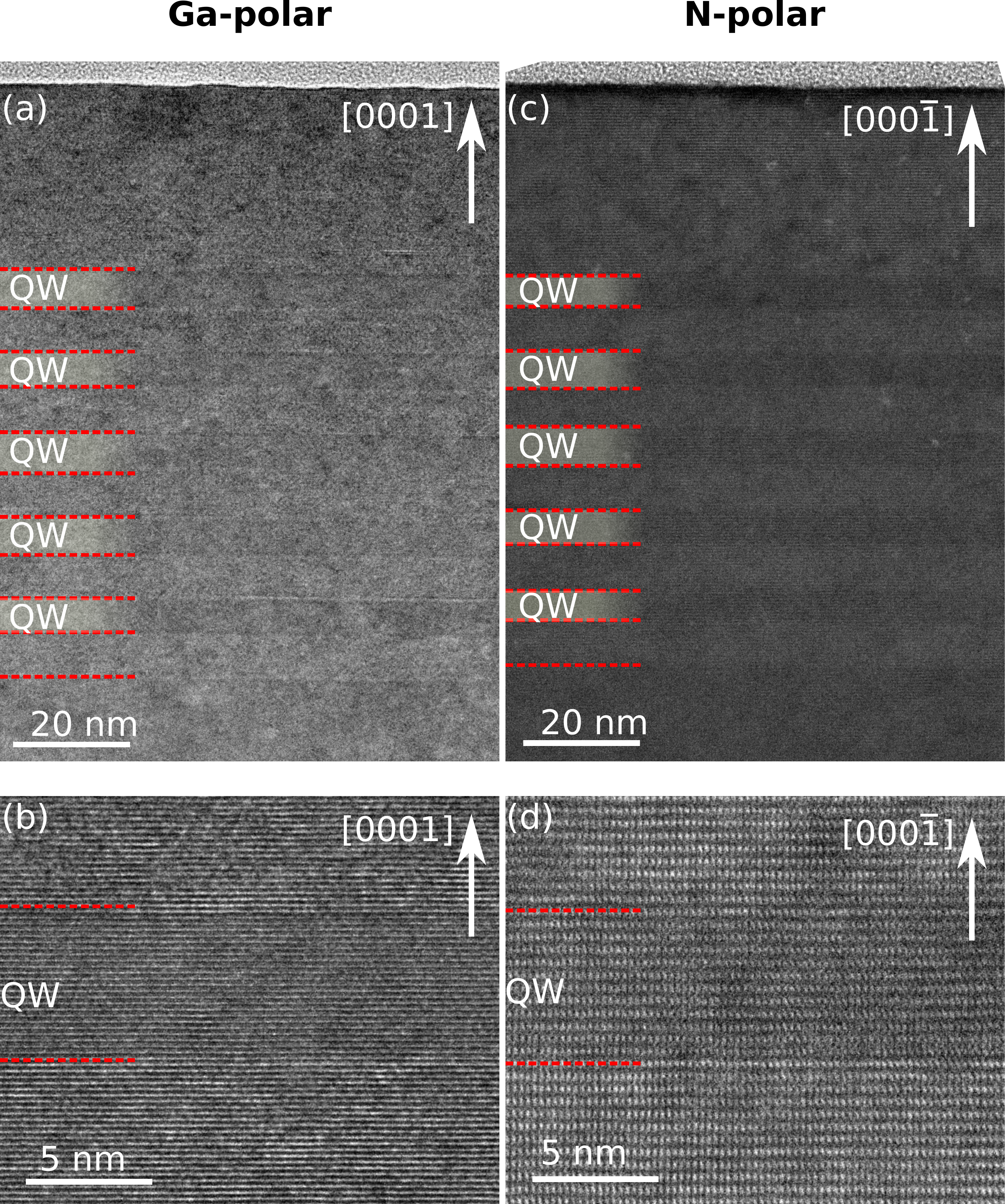}%
\caption{TEM micrographs of Ga-polar (a)--(b) and N-polar (c)--(d) $23/22$~ML GaN/(Al,Ga)N MQWs. (b) and (d) are magnified micrographs over one quantum well of the stack. The dotted lines are guides to the eye to locate the GaN/(Al,Ga)N interfaces. (a)--(b) are taken in the $[1\bar{1}00]$ zone axis and (c)--(d) are taken in the $[11\bar{2}0]$ zone axis.
\label{fig:TEM}}
\end{figure}

TEM analysis of a pair of Ga- and N-polar $23/22$~ML MQWs is performed to get local insights into the heterostructures. Typical micrographs are shown in Fig.~\ref{fig:TEM}. The low amount of Al in the (Al,Ga)N barriers leads to a weak contrast but the full heterostructure remains visible for both samples. The QWs are seen to be all equivalent for each sample and with widths in full agreement with the X-ray predictions. No extended defects such as stacking faults, inversion domains or dislocations can be seen. The limited contrast between the GaN and the (Al,Ga)N layers prevents a quantitative analysis of the interface roughness. For comparison purpose, Sarigiannidou \etal{Sarigiannidou2006} have reported by TEM chemically abrupt interfaces at the monolayer scale for both Ga- and N-polar GaN/AlN MQWs grown at $720$\degC{} by PA-MBE without growth interruptions. The absence of intermixing was also inferred by atom probe tomography \cite{Rigutti2016,Bonef2016}, with the exception of the early work of Mazumder \etal{Mazumder2012}. Therefore, no drastic increase in the interface roughness for N-polar MQWs compared to Ga-polar MQWs should be expected here.  

The combination of AFM, X-ray diffraction and TEM analysis evidences an excellent and equivalent structural quality for the pairs of Ga- and N-polar GaN/(Al,Ga)N MQWs. The absence of extended defects allows us a direct investigation of the nonradiative processes mediated by point defects within the heterostructure. 

\subsection{Impurity concentration}\label{subsec:pointdefects}
\begin{figure}
\includegraphics[width = \linewidth]{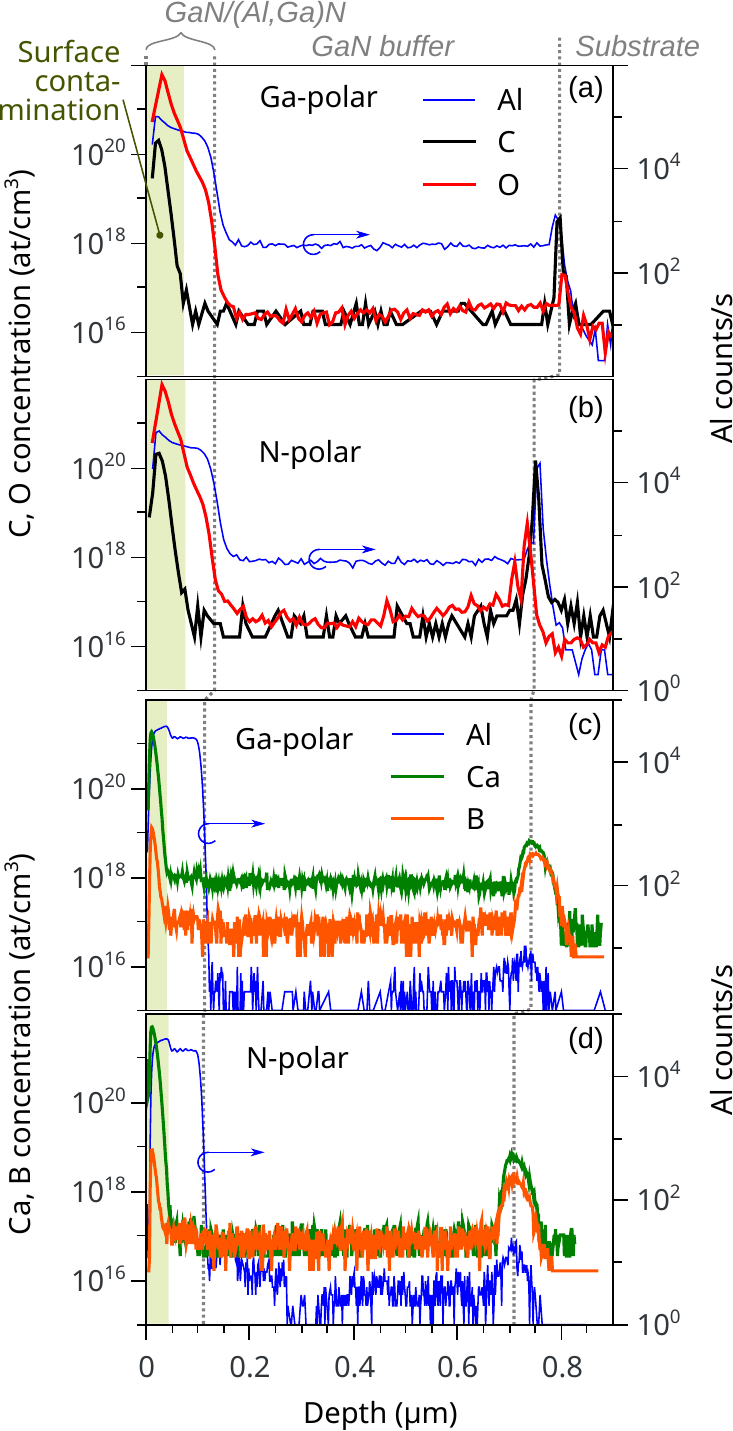}%
\caption{SIMS analysis performed on the Ga- and N-polar $10$~ML GaN/(Al,Ga)N MQWs. (a)-(b) O and C elemental depth profiles acquired using a primary Cs$^+$ ion source; (c)-(d) Ca and B elemental depth profiles acquired using a primary O$_2^+$ ion source. The Al signal is used as a marker and is not quantified.
\label{fig:SIMs}}
\end{figure}

Elemental depth profiles using SIMS are used to measure and compare the concentration of C, O, B and Ca impurities incorporated in the pair of Ga- and N-polar $10$~ML MQWs. The results are shown in Fig.~\ref{fig:SIMs}. The Al signal is used to locate the substrate, the GaN buffer and the actual GaN/(Al,Ga)N heterostructure along the depth profile. The parasitic incorporation of Al in the GaN buffer is attributed to the high beam equivalent pressure of N$_2$, dragging atoms from transversing molecular beams to the substrate analogous to the As drag effect \cite{Wasilewski1997}. For all impurities, a surface peak is observed that corresponds to surface contamination and is thus further neglected.

C, O and B, which are the contaminants typically considered for GaN grown by PA-MBE, are seen with very low levels, down to $2$--$3 \times 10^{16}$\,cm$^{-3}$ for C and O, and down to $7$--$8 \times 10^{16}$\,cm$^{-3}$ for B in the GaN buffer, which is at the very bottom range for GaN grown by MBE \cite{Grandjean1999,Li2000,Kim2001,Ng2002,Storm2012,Diez2020}. Interestingly, the N-polar layer exhibits the same amount of C and O as the Ga-polar one, which is in contrast with some reports dealing with MBE growth \cite{Sumiya2000,Ng2002,Storm2012,Cheze2013b} or HVPE growth \cite{Tuomisto2005} but agrees with some others \cite{Li2000,Okamura2014,Diez2020}. By comparing the substrates and methods used in these different works, it appears that growing the Ga- and N-polar samples in different runs instead of a single one is crucial to obtain equally low impurity incorporation in both polarities. The optimum growth conditions for Ga- and N-polar GaN thus differ. In a more general manner, it follows from these works that impurity incorporation can be readily decreased to very low levels independent of the polarity by minimizing the concentration of extended defects \cite{Popovici1997}, by growing at optimized conditions close to stoichiometry and by minimizing the residues on the substrates surface prior to growth \cite{Storm2016}.
A larger concentration of O and C is observed at the regrowth interface of the N-polar sample compared to the Ga-polar one, which agrees with the usual statement that the GaN$(000\bar{1})$ surface has a thicker native oxide compared to the GaN$(0001)$ surface \cite{Ferguson2015} owing to a larger reactivity of the bare surface toward O \cite{Zywietz1999}. 

If not complexed to other point defects, C, O and B are not expected to act as nonradiative centers in nitrides and can be further disregarded \cite{Mattila1997,Ramos2002}. In contrast, following the work of Young \etal{Young2016}, Ca appears as a common impurity in MBE grown nitrides potentially acting as a strong nonradiative recombination center \cite{Shen2017}. Ca is expected to stem from the substrate surface and to segregate on the growth front where it gradually incorporates in the layer, at higher rates for lower growth temperatures. Fairly high amounts of Ca are detected in the Ga- and N-polar GaN buffers of the investigated samples, with respective concentrations of $8\times 10^{17}$ and $8\times 10^{16}$\,cm$^{-3}$. We recall that these layers were grown at about $700$ and $715$\degC{}, respectively. Ch\`eze \etal{Cheze2018} have also reported a slightly larger Ca concentration of $3$ -- $5 \times 10^{17}$\,cm$^{-3}$ for a N-polar GaN buffer grown by PA-MBE at $730$\degC{} whereas Young \etal{Young2016} found only $8 \times 10^{15}$\,cm$^{-3}$ of Ca impurities in a Ga-polar GaN buffer grown by NH$_3$-MBE at $820$\degC{}. Hence, in addition to the growth temperature \cite{Young2016}, the Ca incorporation rate appears to be drastically affected by the MBE growth method and the crystal orientation. Here, unlike in Ref. [\onlinecite{Young2016}], the reservoir of Ca at the growth front does not deplete after growth of several hundreds of nm of GaN with a Ca concentration above $5\times 10^{17}$\,cm$^{-3}$. This finding indicates a comparatively large amount of Ca initially present at the substrate surface prior the growth ($>0.1$\,ML), which likely originates from the chemomechanical polishing of the freestanding GaN substrates used here and also in the study of Ch\`eze \etal{Cheze2018}.

In the GaN/(Al,Ga)N heterostructures, the impurity concentration is seen to increase by $25\%$, as compared to the GaN buffers for Ca and B atoms. Note that the large increase of the signal close to the surface corresponds to surface contamination. In the case of O, the large surface related signal prevents an unambiguous determination of the O content in the heterostructure. It is likely the presence of Al which triggers the observed increased levels of Ca and B impurities within the heterostructures, as already reported for O incorporation during growth of (Al,Ga)N \cite{Kim2001} or III-As \cite{Achtnich1987,Achtnich1989}.

In conclusion, with the exception of Ca, equally low concentration of impurities are found in the pair of Ga- and N-polar samples. So far, Ca has been reported to act as a nonradiative center for the (In,Ga)N/GaN system, but the question remains open in the case of GaN/(Al,Ga)N.

\subsection{Low temperature photoluminescence}\label{subsec:10KRT}
\begin{figure*}[t]
\includegraphics[width = \linewidth]{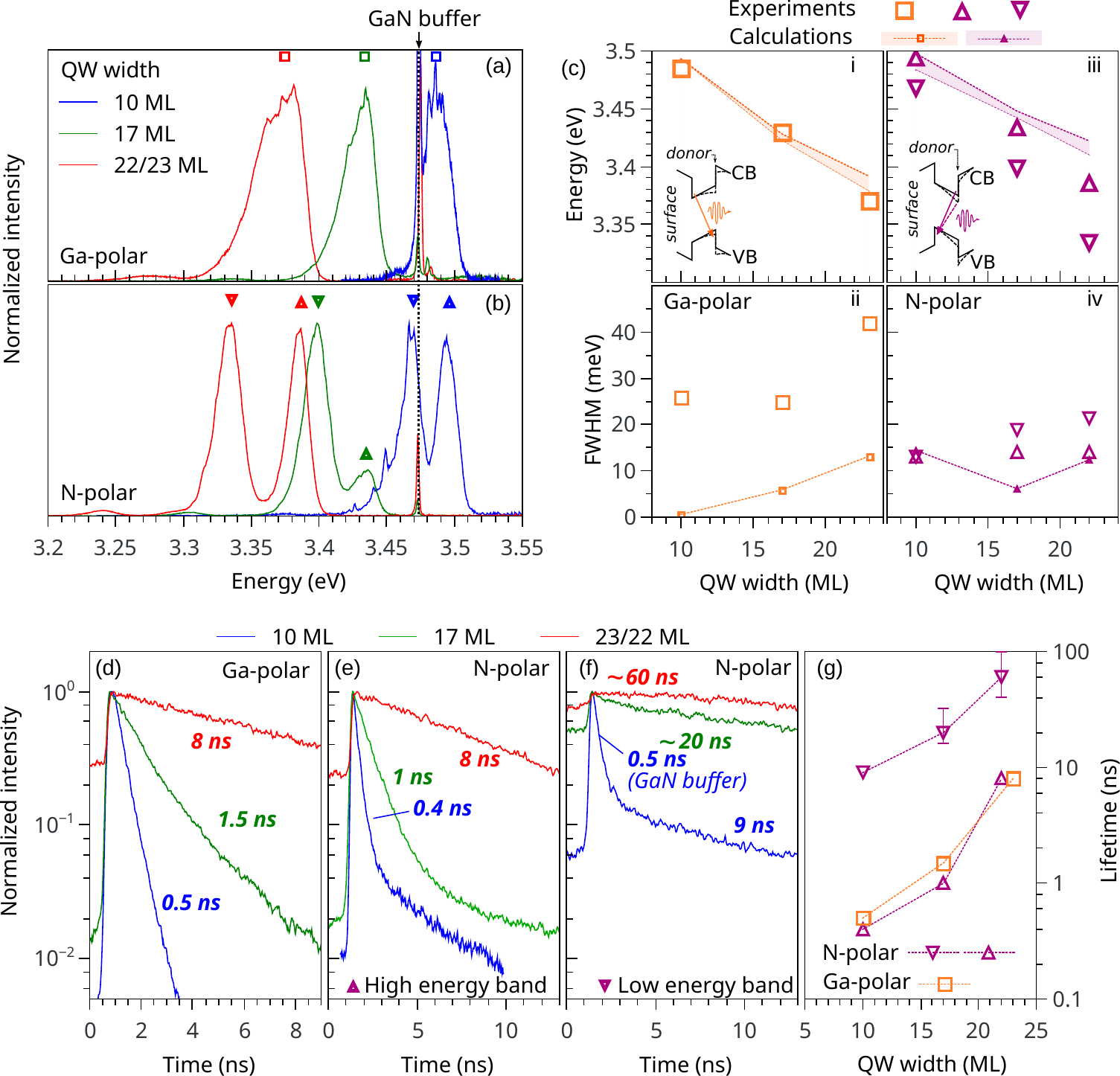}%
\caption{$10$~K PL of the three pairs of GaN(Al,Ga)N MQWs. (a)-(b) Normalized cw-PL spectra acquired with an excitation density of $600$~W/cm$^{2}$ at $325$~nm. (c) Energy and FWHM of the cw-PL band stemming from the MQWs. Dotted lines are results from the 1D Schr\"odinger-Poisson calculations. The shaded areas in i and ii depict the calculated spread in the electron-hole energy separation between the $5$ QWs of the heterostructure. The dotted (plain) band profiles shown in the insets in i and iii schematically depict the modification of the potential profile in the presence (absence) of an ionized donor located at the bottom interface of the QW. CB and VB refer to the conduction and valence band respectively.  (d)--(f) PL transients of the three pairs of GaN/(Al,Ga)N MQWs. (g) Plot of the extracted lifetimes versus the QW width.
\label{fig:LowT}}
\end{figure*}

The cw-PL spectra acquired at $10$~K for the series of Ga- and N-polar MQWs are shown in Fig.~\ref{fig:LowT}(a)--\ref{fig:LowT}(c), along with the extracted energy and width of the observed PL bands. Besides the GaN buffer signal at $3.47$~eV, the MQW PL appears as a broad band accompanied by several phonon replica of lower intensity. Fine structures are visible in each of the bands and are assigned to the localization of holes at the GaN/(Al,Ga)N interface of the MQWs \cite{Lefebvre1998,Grandjean2000,Gallart2000b,Chwalisz2005,Rigutti2016}. Only a weak PL signal related to the (Al,Ga)N barriers is observed at $3.65$~eV (not shown here). Remarkably, it can be seen that the Ga-polar samples exhibit a single PL band whereas the N-polar samples systematically exhibit a doublet. The splitting between the high- and low-energy bands increases from $25$ to $51$~meV, from the narrower to the wider MQWs. The laterally spatially uniform origin of the doublet was confirmed by low temperature monochromatic cathodoluminescence maps of the N-polar $22$~ML MQW surface (not shown here). 

Transients of the PL bands of the various MQWs are plotted in Fig.~\ref{fig:LowT}(d)--\ref{fig:LowT}(f). The lifetimes are extracted by fitting monoexponential decays, with the exception of the low-energy band of the N-polar $10$~ML MQW, for which a biexponential decay is used to account for the overlapping signal from the GaN buffer (fast component with a typical decay time of $0.4$--$0.6$\,ns). Since the acquisition interval was too short to capture the entire decay of the low-energy band of the N-polar MQWs [Fig.~\ref{fig:LowT}(f)], the corresponding error bars are larger than for the high-energy band. The partial spectral overlap between the high- and low-energy bands of the N-polar samples is responsible for the slow component showing up in the transient of the high-energy band of the N-polar $10$~ML MQWs. This slow component is neglected in the following. The extracted lifetimes are plotted in Fig.~\ref{fig:LowT}(g) and show a monotonic increase for wider MQWs. The increase agrees with the decrease in the electron and hole wavefunction overlap due to the internal electric field \cite{Bretagnon2006} and thus shows the dominance of radiative recombination.

In order to identify the contribution of the QCSE in the MQWs PL energy and width, 1D Schr\"odinger-Poisson calculations for the Ga- and N-polar MQWs are performed. The structural parameters of Table~\ref{table:sample_dimensions} and a residual donor concentration of $5 \times 10^{16}$~cm$^{-3}$ are used to simulate the heterostructures. Half of the theoretical value of the polarization coefficients are used since they provide a better agreement with experiments reported so far \cite{Dreyer2016}. The Fermi level energy at the sample surface is set to $0.8$\,eV below the conduction band, as reported for as-grown Ga-polar Al$_{0.1}$Ga$_{0.9}$N \cite{Reddy2015}. The same value is used for Ga- and N-polar heterostructures since there are no reports for this quantity for N-polar (Al,Ga)N. The calculation outputs the band profile and the internal electric field of the heterostructure. Exemplary results are shown in the appendix (Fig.\,\ref{fig:MQS_calc}) for the $23/22$\,ML GaN/(Al,Ga)N MQWs. In addition, the calculations deliver single-particle transitions for each of the QWs of the heterostructure, which is used as an estimate for the PL energy of the MQWs (excitonic effects are ignored). Each QWs of the same heterostructure have different environments (\textit{e.g.}, distance to the surface) which leads to a different amplitude of their internal electric field. The differences in the single-particle transition energies for the $5$ QWs of each samples are then used as an estimate of the full-width at half-maximum of the MQW's PL signal. 

For the Ga-polar samples, an excellent agreement is found between the calculated and experimental PL energies, as shown in Fig.~\ref{fig:LowT}(c). Only the calculated FWHMs are underestimated, which can be attributed to the alloy disorder in the (Al,Ga)N barriers, to the random positions of ionized dopants \cite{Morel2001} and to monolayer fluctuations in the QW widths \cite{Grandjean2000,Natali2009}. According to the calculations, the internal electric field within the GaN QWs amounts to ($280 \pm 50$)~kV/cm.  Since the recombination energy and rate of the Ga-polar MQWs at $10$\,K can be entirely ascribed to the presence of the internal electric field, we then conclude that the exciton decay within the MQWs is predominantly radiative, as typically reported for high quality Ga-polar GaN/(Al,Ga)N heterostructures \cite{Smith1996,Zheng1999}. 

For the N-polar samples, the calculated PL energies match well the high-energy band, but largely overestimate the low-energy band. The calculated FWHMs are also underestimating the experimental values, likely for the same reasons invoked for the Ga-polar samples. The internal electric field amounts to ($170 \pm40$)~kV/cm, which is lower than for the Ga-polar QWs. This reduction is attributed to the surface band bending since in the N-polar (Ga-polar) case, its electric field is opposed (aligned) to the one generated by the spontaneous and piezoelectric polarizations. In analogy to the Ga-polar case, we attribute the high-energy PL band of the N-polar MQWs to a predominantly radiative excitonic recombination dominated by the QCSE. 

Concerning the low-energy PL band, we consider several different scenarios. First, the width of GaN/(Al,Ga)N QWs is commonly observed to fluctuate by $\pm2$\,ML, potentially giving rise to multiple excitonic features in the PL spectra \cite{Gallart2001,Haratizadeh2007,Natali2009}. Yet, according to Fig.\,\ref{fig:LowT}(c), a net increase in the QW width of $+4$~ML is required to account for the $50$~meV splitting between the PL band of the wider N-polar MQWs. Our XRD and TEM data (cf.\ Figs.~\ref{fig:xray} and \ref{fig:TEM}) do not exhibit any evidence for such a large thickness fluctuation. In addition, the PL spectra do not show any signal associated to QW widths of $+1$, $+2$ and $+3$~ML, as would be expected for a GaN/(Al,Ga)N interface with such a degree of roughness. Furthermore, the exceptionally long decay time of the low-energy as compared to the high-energy band could not be explained by recombination of excitons in a QW wider by a few MLs only. 

Second, an even larger splitting between transitions of Al$_{0.7}$Ga$_{0.3}$N/Al$_{0.6}$Ga$_{0.4}$N MQWs grown by MOCVD has been attributed to biexcitons with an extraordinarily large binding energy \cite{Hayakawa_2014}. However, much lower biexciton binding energies have been observed in GaN/(Al,Ga)N QWs with low Al content \cite{Cheregi_2008a}. For the $9\%$ Al content corresponding to our samples, a biexciton binding energy of only 2\,meV is expected for a well width of 10\,ML, and even lower values are extrapolated for thicker wells due to the impact of the internal electric fieds in QWs with a width exceeding the exciton Bohr radius \cite{Cheregi_2008b}. In addition, we have performed excitation-density-dependent measurements (not shown here) that reveal the low-energy line to saturate for higher excitation densities, suggesting that this line is associated with an impurity rather than being of intrinsic origin.   

Third, bound excitons are an excitonic complex analogous to the biexciton, but tied to an impurity. The donor-bound exciton usually dominates the PL spectra of bulklike GaN and has a binding energy of 6--7\,meV. He \etal{He2015} have attributed a splitting of about 60 (90)\,meV in the PL spectra of Al$_{0.5}$Ga$_{0.5}$N/Al$_{0.35}$Ga$_{0.65}$N MQWs grown by MOCVD to the recombination of Si (O) donor-bound excitons. However, such a large splitting is inconceivable in a GaN QW. The maximum donor ionization energy in GaN/Al$_{0.15}$Ga$_{0.85}$N QWs has been calculated to be 60\,meV \cite{Morel2001}, which translates into a binding energy of the donor-bound exciton of 12\,meV, much too small to account for the splitting experimentally observed. 

The donor ionization energy itself, however, is close to this splitting. Last, we thus consider the radiative recombination between a shallow donor (such as O) and a free hole. Free-to-bound recombination is commonly observed in doped GaAs/(Al,Ga)As heterostructures \cite{Balchin1995} but has been scarcely considered in GaN/(Al,Ga)N heterostructures. However, in GaN/(Al,Ga)N-based QWs with a width exceeding the exciton Bohr radius, such as the samples under investigation in the present work, excitonic effects are suppressed due to charge separation \cite{Cheregi_2008b}, and nonexcitonic recombination (such as band-to-band and free-to bound transitions) may become dominant. Free-to-bound transitions are also known to have a significantly longer lifetime than band-to-band transitions \cite{Balchin1995}. A donor-to-free hole transition is thus consistent with both, the energy and the lifetime of the low-energy line of our QWs.

Furthermore, interpreting the low-energy PL line of the N-polar MQWs as a donor-to-free hole transition offers not only an understanding of the spectral and temporal characteristic of this transition, but also for the absence of the splitting in the Ga-polar MQWs. In fact, the donor binding energy depends on the position of the donor relative to the QW interfaces, on the width of the QW, and, most importantly, on the polarity of the heterostructure \cite{Morel2001}. The influence of polarity results from the difference in overlap between the electron and the ionized donor, which strongly depends on whether the electron resides at the donor interface or at the opposite one [this situation is exemplified in the band profiles shown as insets in Figs.\,\ref{fig:LowT}(c)i and \ref{fig:LowT}(c)iii]. In particular, for a donor located at the bottom interface of a $22$\,ML wide N-polar (Ga-polar) GaN/Al$_{0.15}$Ga$_{0.85}$N QW, a binding energy of $50$\,meV ($20$\,meV) is obtained \cite{Morel2001}. In other words, we would expect the free-to-bound transition to create a distinct doublet for the N-polar samples, but for the Ga-polar sample, the free-to-bound transition would be expected to  spectrally overlap with the band-to-band transition.

The most abundant shallow donor in our structures is O. The preferential incorporation of O at the bottom interface (the “inverted” interface) is well documented for GaAs/(Al,Ga)As MQWs grown by MBE, and is attributed to the formation of an AlO$_x$ floating layer during (Al,Ga)As growth \cite{Achtnich1989,Chand_1993}. It is very plausible that an analogous mechanism to preferentially incorporate O at the bottom interface is at work for GaN/(Al,Ga)N QWs as well. Overall, this hypothesis is the only one consistent with all experimental findings: the magnitude of the splitting for the N-polar samples, the long lifetime of the low-energy line, its saturation for high excitation density, and the absence of a detectable splitting for the Ga-polar samples.


\subsection{Room-temperature photoluminescence}\label{subsec:RTPL}
\begin{figure*}[h]
\includegraphics[width = .7\linewidth]{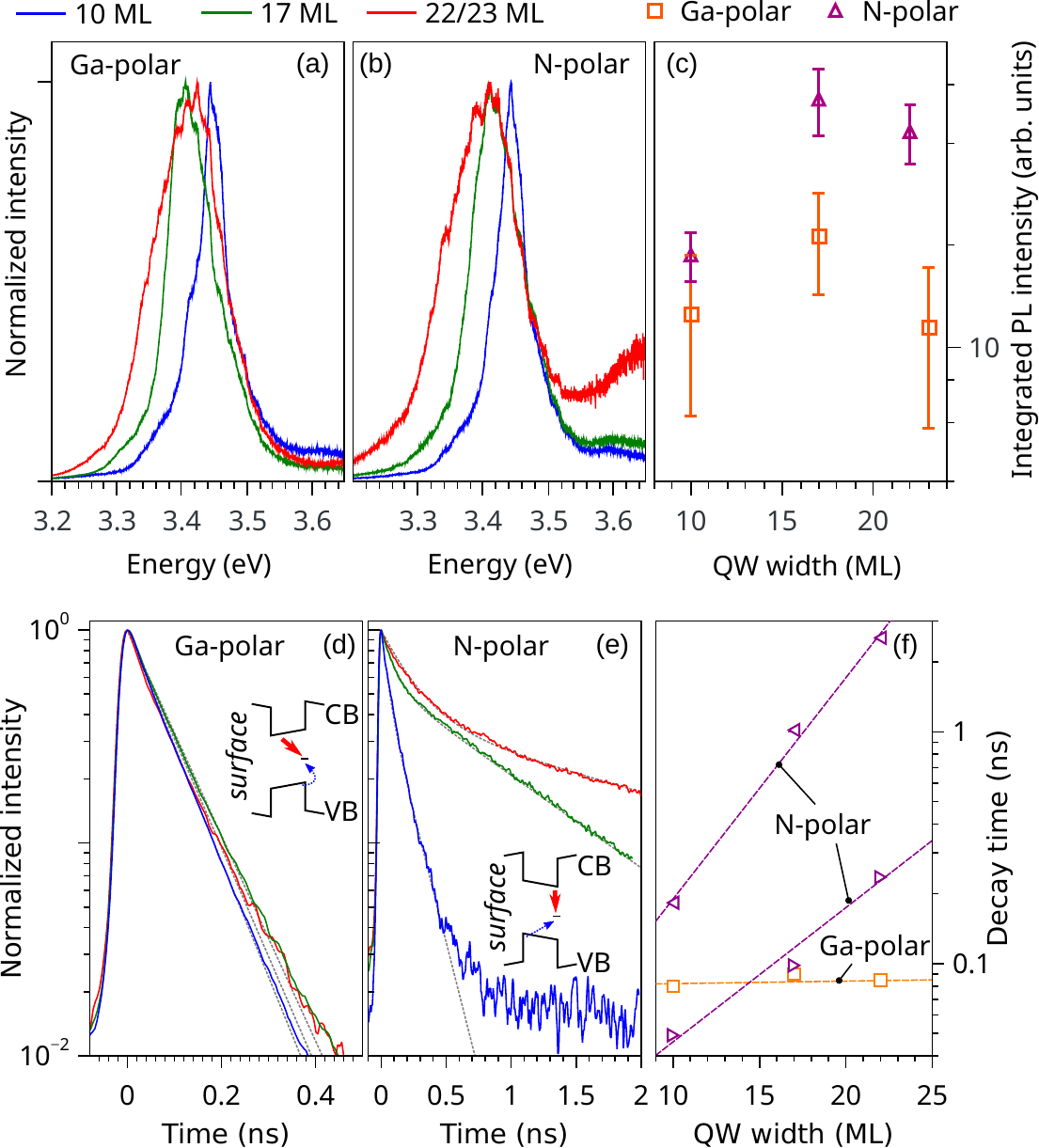}%
\caption{Room-temperature PL of the three pairs of GaN/(Al,Ga)N MQWs. (a)-(b) Normalized cw-PL spectra acquired with excitation at $244$~nm with a power density of $50$~kW/cm$^{2}$. (c) Integrated intensity of the GaN/(Al,Ga)N MQWs cw-PL between $3.15$ and $3.6$~eV. Error bars account for the dispersion at the scale of the sample. (d)-(e) PL transients of the GaN/(Al,Ga)N MQWs. Dotted lines show the result of monoexponential and biexponential fits for the Ga- and N-polar samples, respectively. The insets depict a nonradiative decay path involving a hole trap state located at the bottom QW interface. Blue dotted (red plain) arrows refer to the decay of holes (electrons). (f) Non-radiative decay time extracted from the fits plotted as function of the MQW widths. Dashed lines are guides to the eye.
\label{fig:RT-PL}}
\end{figure*}

Fig.~\ref{fig:RT-PL}(a)--\ref{fig:RT-PL}(b) shows the normalized cw-PL spectra of the three pairs of samples acquired at room-temperature. The PL signal is confirmed to originate from the QWs by comparison with the PL spectra of a sample free of QWs (not shown). The average intensity of the MQWs' cw-PL is plotted in Fig.~\ref{fig:RT-PL}(c) and exhibits a non-monotonous dependence on the width of the MQWs, which is more pronounced for the N-polar heterostructures. Interestingly, N-polar samples exhibit a higher luminescence intensity than Ga-polar ones, up to a factor $3$ for the thickest QWs of the series. 

Such enhanced radiative efficiency of the N-polar samples is further confirmed by observing the PL transients acquired at room-temperature and as shown in Fig.~\ref{fig:RT-PL}(d)--\ref{fig:RT-PL}(e) for both polarities. The transients are fitted by monoexponential and biexponential decays for the Ga- and N-polar samples, respectively. The obtained characteristic times are plotted in Fig.~\ref{fig:RT-PL}(f) as function of the QW width. The observed decay times are systematically shorter than the one acquired at $10$~K, which indicates that electron-hole recombination is predominantly nonradiative at room-temperature \cite{Andreani1991}, a phenomenon typically observed for such type of heterostructures \cite{Smith1996,Zheng1999}. For the Ga-polar MQWs, the decay time of $0.1$~ns has essentially no dependence on the width of the individual QWs. In contrast, the decay times in the N-polar MQWs lengthen for larger wells, reaching up to $2$~ns. This behavior is reminiscent of nonradiative recombination taking place at the interfaces of the QWs. The usual description of this decay channel by a characteristic time $\tau_{nr} = d/2S$ \cite{Nelson1978,Wolford1994}, with $d$ the QW width and $S$ the interface recombination velocity, is not applicable here since the out-of-plane location of carriers is primarily determined by the internal electric field and not by carrier diffusion. Instead, an exponential dependence of the decay time on the QW width is observed [$\tau_{nr} \propto \exp(d/L)$], with a characteristic length $L = 1$--$2$\,nm, suggesting that the lifetime is controlled by the wavefunction overlap of spatially separated states.

In the following, we discuss the case where the point defects responsible for efficient nonradiative recombination in the Ga- and N-polar GaN/(Al,Ga)N QWs are exclusively located at the bottom QW interface. The recombination scheme proposed in this scenario is schematized in the insets of Figs.\,\ref{fig:RT-PL}(d) and \ref{fig:RT-PL}(e). We further assume that the nonradiative recombination rate is limited by the trapping of minority carriers which are holes. As a matter of fact, in the Ga-polar samples, the excess of electrons is indeed substantial since the Fermi level intersects the bottom of the conduction band in each QWs (see Fig.\,\ref{fig:MQS_calc} in the appendix). Under laser exposure, photoexcited holes in N-polar (Ga-polar) QWs are driven to the top (bottom) interface by the internal electric field. For the N-polar samples, the capture rate of holes by the traps located at the bottom interface should thus decrease exponentially with the well width due to the vanishing wavefunction overlap. In contrast, no dependence of the hole capture rate on the well width is expected for the Ga-polar samples. This simple consideration on the hole capture rate can thus account for the experimental observation that only N-polar samples exhibit a well width dependent nonradiative decay time. 

Large densities of interface hole traps have been inferred in N-polar GaN/(Al,Ga)N heterostructures either grown by MOCVD \cite{Prozheeva2020} or MBE \cite{Mohanty2020}. As a likely candidate, the N vacancy (V$_N$) was proposed, which forms two acceptor-like states in the bandgap of bulk GaN \cite{Lyons2017}. Ga-polar (In,Ga)N/GaN heterostructures grown by MOCVD are also expected to incorporate high densities of V$_N$ \cite{Chen2021}. For MOCVD growth, the current understanding is that V$_N$ form during high temperature GaN growth where thermal GaN decomposition is non negligible ($>900$\degC{}), segregate at the surface even at low growth temperatures and eventually incorporate in the bulk at the onset of (In,Ga)N growth \cite{Akasaka2004,Armstrong2015,Haller2017,Haller2018,Chen2021}. Similarly, in N-polar GaN/(Al,Ga)N QWs grown by MOCVD, the presumed V$_N$ form at the onset of the ternary alloy growth (at the top QW interface) \cite{Prozheeva2020}. However, for MBE grown GaN/(Al,Ga)N QWs, the formation mechanism for V$_N$ may differ. Another potential candidate for the trap state is the impurity Ca that was reported to act as a deep acceptor mediating efficient nonradiative recombination in (In,Ga)N QWs \cite{Young2016,Shen2017}. In our case, the N-polar samples contain ten times less Ca as the Ga-polar samples, which correlates with their enhanced radiative efficiency. Yet, \emph{ab initio} calculations of bulk (In,Ga)N doped with Ca show that the electron capture coefficient of the deep state strongly decreases when increasing the band gap, becoming essentially zero above $2.8$\,eV \cite{Shen2017}.

We thus propose that the nonradiative recombination observed at room-temperature for the GaN/(Al,Ga)N MQWs results from Shockley-Read-Hall recombination on point defects (presumably Ca or V$_N$) located at the bottom GaN QW interface. For N-polar heterostructures, this nonradiative process is significantly affected by the internal electric field. By increasing the QW width, the nonradiative recombination time is not constant but increases due to the reduction in the overlap between the electron and hole wavefunctions. Overall, the internal quantum efficiency is essentially given by the ratio between the nonradiative and radiative lifetime. For the sample series under scrutiny, this ratio appears not to have a monotonous dependence on the QW width since the maximum PL intensity is observed for a QW width of $17$\,ML. 

\section{SUMMARY AND CONCLUSIONS}
Structurally equivalent Ga- and N-polar GaN/(Al,Ga)N MQWs are grown on freestanding GaN by PA-MBE. Under optimized growth conditions, no more extended defects are found in the heterostructures than originally present in the substrates. Low concentration of impurities are found in samples of opposite polarity, with the exception of Ca, which was measured with a ten times higher concentration ($8 \times 10^{17}$~cm$^{-3}$) in the Ga-polar MQWs compared to the N-polar MQWs. At $10$~K, the exciton decay within the MQWs is deduced to be predominantly radiative for all samples and the PL properties are entirely characterized by the QCSE. For the N-polar MQWs, an additional low-energy line is observed and ascribed to a free-to-bound recombination involving a shallow donor, likely O, aggregated at the QW bottom interface. At room-temperature, the electron-hole pair recombination is governed by nonradiative processes for all samples. The decay rate in N-polar MQWs shows a dependence on the QW width and is lower than in Ga-polar MQWs. A nonradiative decay mediated by Shockley-Read-Hall recombination on deep states located at the bottom QW interface is proposed to account for these observations. Remarkably, the presence of internal electric fields is seen as beneficial in N-polar GaN/(Al,Ga)N MQWs for slowing down the nonradiative recombination. Nevertheless, further enhancement of the internal quantum efficiency of GaN/(Al,Ga)N heterostructures grown by MBE will necessarily require to minimize the density of point defects. Two approaches currently under developments are the growth of dedicated buffer layers to bury point defects \cite{Chen2021,Chaudhuri2021} and the use of different growth conditions \cite{Wang2021,Tatarczack2021}.

\appendix*
\section{Calculated band profiles}

\begin{figure}[b]
	\includegraphics[width = \linewidth]{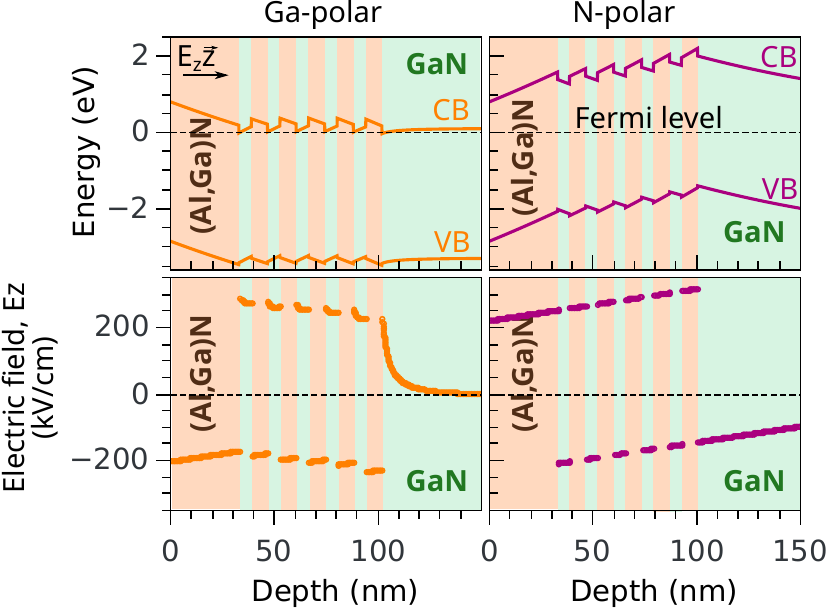}%
\caption{Calculated conduction band (CB) and valence band (VB) structures and internal electric field for the Ga- and N-polar $23/22$\,ML GaN/(Al,Ga)N MQWs.
\label{fig:MQS_calc}}
\end{figure}

The band structure and electric field distribution obtained by Schr\"odinger-Poisson calculations of the thicker MQWs are plotted in Fig.\,\ref{fig:MQS_calc}. Clear differences can be observed between the Ga- and N-polar cases. In the first case, the Fermi level is located close to the conduction band minimum and the average internal electric field amounts to $280$\,kV/cm in the QWs. In the second case, the Fermi level is close to the mid-gap position and the average internal electric field amounts to $200$\,kV/cm within the QWs. In both cases, the electric field differs between each of the QWs of a MQW sample, with variations up to $15\%$.

%
\begin{acknowledgments}
 We are indebted to Katrin Morgenroth for her support during the preparation of the samples as well as for her dedicated maintenance of the MBE system together with Carsten Stemmler and Hans-Peter Sch\"onherr. We thank Vladimir Kaganer for stimulating discussions concerning X-ray characterization and Manfred Ramsteiner for a critical reading of the manuscript. Funding from both the Bundesministerium für Bildung und Forschung through project FKZ:13N13662 and the Leibniz-Gemeinschaft under Grant SAW-2013-PDI-2 is gratefully acknowledged. Sergio Fernández-Garrido acknowledges the partial financial support received through the Spanish program Ram\'on y Cajal (co-financed by the European Social Fund) under grant RYC-2016-19509 from the former Ministerio de ciencia, innovaci\'on y universidades.
\end{acknowledgments}


%

\end{document}